\documentclass[12pt]{article}
\usepackage{graphicx}%
\newcommand{\bda}{\begin{\displaymath}\begin{array}{rl}}
\newcommand{\eda}{\end{array}\end{displaymath}}
\newcommand{\be}{\begin{equation}}
\newcommand{\ee}{\end{equation}}
\newcommand{\bdm}{\begin{displaymath}}
\newcommand{\edm}{\end{displaymath}}
\newcommand{\bea}{\begin{eqnarray}}
\newcommand{\eea}{\end{eqnarray}}

\begin{document}
\vspace{5em}
\begin{center}
{\bf \Large Finding the sigma pole by analytic extrapolation of $\pi\pi$ scattering  data} 

\vspace{2em}
 Irinel Caprini,

\vspace{1em}
 National Institute of Physics and Nuclear Engineering\\
POB MG 6, Bucharest, R-077125 Romania

\end{center}

\vskip1cm
\begin{abstract}
 We investigate the determination of the $\sigma$ pole  from  $\pi\pi$ scattering data below the $K\bar{K}$ threshold, including the new  precise results  obtained from  $K_{e4}$ decay by NA48/2 Collaboration. We discuss also the experimental status of the threshold parameters $a_0^0$ and $b_0^0$ and the phase shift $\delta_0^0$. In order to reduce the theoretical bias, we use a large class of analytic parametrizations of the isoscalar $S$-wave, based on expansions in powers of conformal variables. The  $\sigma$ pole obtained with this method is consistent with the prediction based on
 ChPT and Roy equations. However, the  theoretical uncertainties are  now larger, reflecting the sensitivity of the pole position to the specific parametrizations valid in the physical region. We conclude that Roy equations offer the most precise method for the determination of the $\sigma$ pole from $\pi\pi$ elastic scattering.
\end{abstract}

\hspace{0.38cm} PACS: 13.75.Lb, 14.40.Cs

\section{Introduction}
The determination of the pole associated to the  $\sigma$ resonance  (or $f_0(600)$)  is known  to be a   difficult problem. The pole is situated deep in the complex plane, its influence in the physical region is masked to a certain extent by the nearby Adler zero  and, until recently,  the experimental data on  $\pi\pi$ scattering
  at low energies were quite poor.  This explains why the values reported by PDG \cite{PDG2006} for the mass and width of  $\sigma$  cover a very large interval.

During the last years,  Chiral Perturbation Theory (ChPT) and Roy equations led to an accurate description of $\pi\pi$ scatering at low energies  \cite{ACGL, CGL}. In  particular, the scattering length $a_0^0$ and the effective range parameter  $b_0^0$ of the isoscalar $S$-wave given in \cite{CGL}:
\be\label{abCGL}
 a_0^0=0.220 \pm 0.005, \quad\quad b_0^0=0.276 \pm 0.006\,,
\ee
have remarquably small uncertainties. 

 The formalism based on Roy equations  was shown recently  \cite{CCL} to control also the analytic extrapolation of the $\pi\pi$ amplitude in the complex plane, leading to precise values for the mass and width of $\sigma$:
\be\label{CCL}
M_\sigma = 441\,^{+16}_{-8}\, {\rm MeV}, \quad \quad \Gamma_\sigma/2= 272\,^{+9}_{-12.5} \, {\rm MeV}.
\ee 
In the standard method  of detecting resonances, the experimental data on the partial wave with the quantum numbers of the resonance  play an important role. Unlike this,
 the prediction (\ref{CCL})  was obtained without using  experimental data on the isoscalar  $S$-wave at low energies: the amplitude  was calculated below 800 MeV, and also in the complex plane, from Roy equations, using experimental input at higher energies and theoretical results on the pion-pion scattering \cite{CGL}. Roy equations provide a very suitable framework in this case, compensating the lack of experimental data on $\pi\pi$ scattering at low energies by theoretical information.

Recently \cite{NA48},  NA48/2 Collaboration  measured the phase shift difference $\delta_0^0-\delta_1^1$ at low energies from  $K_{e4}$ 
 decay,  with a  precision much greater than that of the older experiments
 \cite{Rosselet, Pislak}. This revived the interest in 
the determination of the scattering length $a_0^0$ and the
 pole associated to $\sigma$ by direct analytic extrapolation of the $\pi\pi$ scattering data. In \cite{GMPY}
  the authors propose a
  representation of the isoscalar $S$-wave $t_0^0(s)$ based on an  expansion in powers of a conformal mapping variable. To account for the theoretical
 uncertainties related to analytic continuation, 
 two  parametrizations  were
 considered, the difference between them being interpreted as a systematic theoretical  uncertainty of the method. In the framework discussed in \cite{GMPY},  the mass and width of $\sigma$ are obtained with an accuracy comparable to that quoted in (\ref{CCL}). 

In the present work we focus on the problem of systematic uncertainties within this approach. We note that the class of functions used in  \cite{GMPY}, although based on a convergent expansion, is still quite narrow when the expansion is restricted to a few terms. By enlarging the class of admissible analytic functions used for fitting the data,  the theoretical bias is reduced and 
 a more realistic estimate of the uncertainties in the position of the $\sigma$ pole is obtained. In the present work we apply this idea,  by using  a large sample of analytic parametrizations of the $\pi\pi$ amplitude, suitable at low energies.  A short description of the method and some results were given already in \cite{ICScadron}.

In the next section we discuss several parametrizations of the $\pi\pi$ isoscalar $S$-wave, which satisfy analyticity and elastic unitarity. In the next two sections  we apply these parametrizations for fitting the data on the phase shift $\delta_0^0$: in section 3 we consider only the data from $K_{e4}$ decay, and in section 4 we include data up to the $K\bar{K}$ threshold. From  the  admissible parametrizations of the isoscalar $S$-wave we
 find the threshold parameters $a_0^0$ and $b_0^0$ and the location of the $\sigma$ pole. Our conclusions are summarized in  section 5. 

\section {The isoscalar $S$-wave at low energy}
 We consider the $\pi\pi$ isoscalar $S$-wave $t_0^0(s)$, which is an analytic function in the $s$-plane cut along $s\ge 4  M_\pi^2$ and $s \le 0$. We assume that  $t_0^0(s)$ is the pure strong amplitude, where all the isospin breaking corrections  are neglected. As in \cite{ACGL}-\cite{CCL}, we take for  $M_\pi$ and $M_K$ the masses of the charged pion and charged kaon, respectively.

Neglecting the inelasticity due to the 4$\pi$ channel below 1 GeV,  unitarity implies that the relation
\be \label{unit}
 \mbox{Im} \left[\frac{1}{t_0^0(s+i\epsilon)}\right] = - i \rho(s),\quad\quad\quad \rho(s)=\sqrt{1-  4 M_\pi^2/s},
\ee  
is valid up to  the threshold for $K\bar K$ production, $s= 4 M_K^2$.
From (\ref{unit}) it follows that the function $\psi(s)$, defined by
\be\label{t00}
t_0^0(s)= \frac{1}{\psi(s)- i \rho(s)},
\ee 
is real in the elastic region:
\be\label{imag}
\mbox{Im} \psi(s+i\epsilon) =0,\quad \quad \quad   4 M_\pi^2 \le s < 4 M_K^2,
\ee  
and is related to the phase shift $\delta_0^0(s)$ by:
\be\label{delta00}
 \psi(s)=\rho(s)\, \mbox{cot}\delta_0^0(s).
\ee 
 Since the  amplitudes are analytic functions of real type, Eq. (\ref{imag}) means that 
 $\psi(s)$ has no discontinuity across the elastic unitarity cut.
The definition (\ref{t00}) shows also that $\psi(s)$ has poles at the points where $t_0^0(s)$ has zeros. The amplitude is expected to vanish below threshold at a point $s_A$, related to the so-called  Adler zeros.   ChPT to lowest order predicts $s_A=M_\pi^2/2$. Assuming that $t_0^0(s)$  does not have other  zeros in the complex plane, the product $(s- s_A) \psi(s)$ is analytic
in the $s$-plane cut for  $s \le 0$ and  $s\ge 4 M_K^2$. The effective range expansion amounts to
  expanding the function $\psi(s)$
 in powers of $q^2=(s/4- M_\pi^2)$ around the threshold $q^2=0$, where it is regular.  However, the branch point $s=0$ limits the convergence of this expansion  to the circle $|q^2|< M_\pi^2$.
\subsection {Method of conformal mappings}\label{conf}
 The domain of convergence of a power series can be enlarged by expanding the function in powers of a variable which  conformally maps a part of the holomorphy domain onto the interior of a disk. The use of conformal mappings in particle  physics was first discussed in  \cite{Confmap, CutDeo1}; in the context of the effective range  expansion for partial waves a conformal mapping was used in \cite{CutDeo};  more recently, the method was applied  for the description of exclusive semileptonic  $B$ decays \cite{CaLeNe} and in perturbative QCD \cite{CaFi}.  As shown in \cite{Confmap},  the asymptotic rate of convergence of the series in the physical region is optimal if the amplitude is expanded in powers of the variable which  mapps the entire holomorphy domain onto a disk. Since the disk is the natural convergence domain of the power series, the new expansion will converge in the whole analyticity domain, up to its boundary.

Consider the variable  
\be\label{wa}
 w(s,\alpha )=\frac{\sqrt{s}- \alpha \sqrt{4 M_K^2-s}}{\sqrt{s} + \alpha \sqrt{4 M_K^2-s}},
\ee
where  $\alpha >0$ is arbitrary. The function  $w(s, \alpha)$ transforms the $s$-plane cut along $s\le 0$ and $s\ge 4 M_K^2$ onto the unit disk  
$|w|<1$ in the complex plane $w= w(s,\alpha )$,  such that $w(4 M_K^2,\alpha )=1$ and $w(0, \alpha)=-1$. In  \cite{GMPY}  the authors adopt the expansion 
\be\label{psiYgen}
\psi(s)= \frac{M_\pi^2}{s-s_A} \left[ \frac{2 s_A} {M_\pi\sqrt{s}}+ B_0+B_1 w(s, \alpha) +B_2 w(s, \alpha)^2+\ldots\right]
\ee  
with  the particular choice  $\alpha=1$. In Eq. (\ref{psiYgen}), the first term in parantheses, added to the expansion in powers of  $w(s,\alpha)$, compensates the singularity of $\rho(s)$ at $s=0$ in the denominator of (\ref{t00}),  removing an unphysical singularity  of $t_0^0(s)$ on the real axis  which would appear otherwise.

 A slightly  different form was also used in  \cite{GMPY}:
\be\label{psiY}
\psi(s)= \frac{M_\pi^2}{s-s_A}\frac{\mu_0^2-s}{\mu_0^2} \left[ \frac{2  s_A}{M_\pi\sqrt{s}}+ B_0+B_1 w(s,1 )+B_2 w(s, 1)^2+\ldots \right],
\ee
where the factor $(\mu_0^2-s)$ displays explicitly the energy  where the  phase shift $\delta_0^0$  passes through $\pi/2$, according to (\ref{delta00}).  This factor is useful for fitting narrow resonances  but, as we shall show, it is not suitable for broad resonances like $\sigma$.

 The power expansions in (\ref{psiYgen}) and (\ref{psiY}) converge in the disk $|w|<1$  and, for a large number of terms,  these parametrizations are equivalent. However, when the series are truncated at a finite number of terms, (\ref{psiYgen}) and (\ref{psiY}) lead to different results. This difference is interpreted
  in  \cite{GMPY} as a systematic uncertainty of theoretical nature,  which should be added to the statistical errors. In the present work we develop this idea, presenting  other admissible analytic parametrizations of the amplitude.

A first generalization  is to expand $\psi(s)$ in powers of $w= w(s,\alpha )$, for an arbitrary $\alpha$, as in (\ref{psiYgen}).
 By varying $\alpha$, one changes the point mapped to the origin of the $w$-plane and the position of the intervals where experimental data are available. Some examples  are shown in Fig. 1.  As we shall see, the flexibility offered by the parameter $\alpha$  allows us to describe the peculiar structure of the isoscalar $S$-wave near the inelastic $K\bar K$ threshold.

\begin{figure}[thb]\label{wgen}\begin{center}\vspace{0.3cm}
\includegraphics[width=4.5cm]{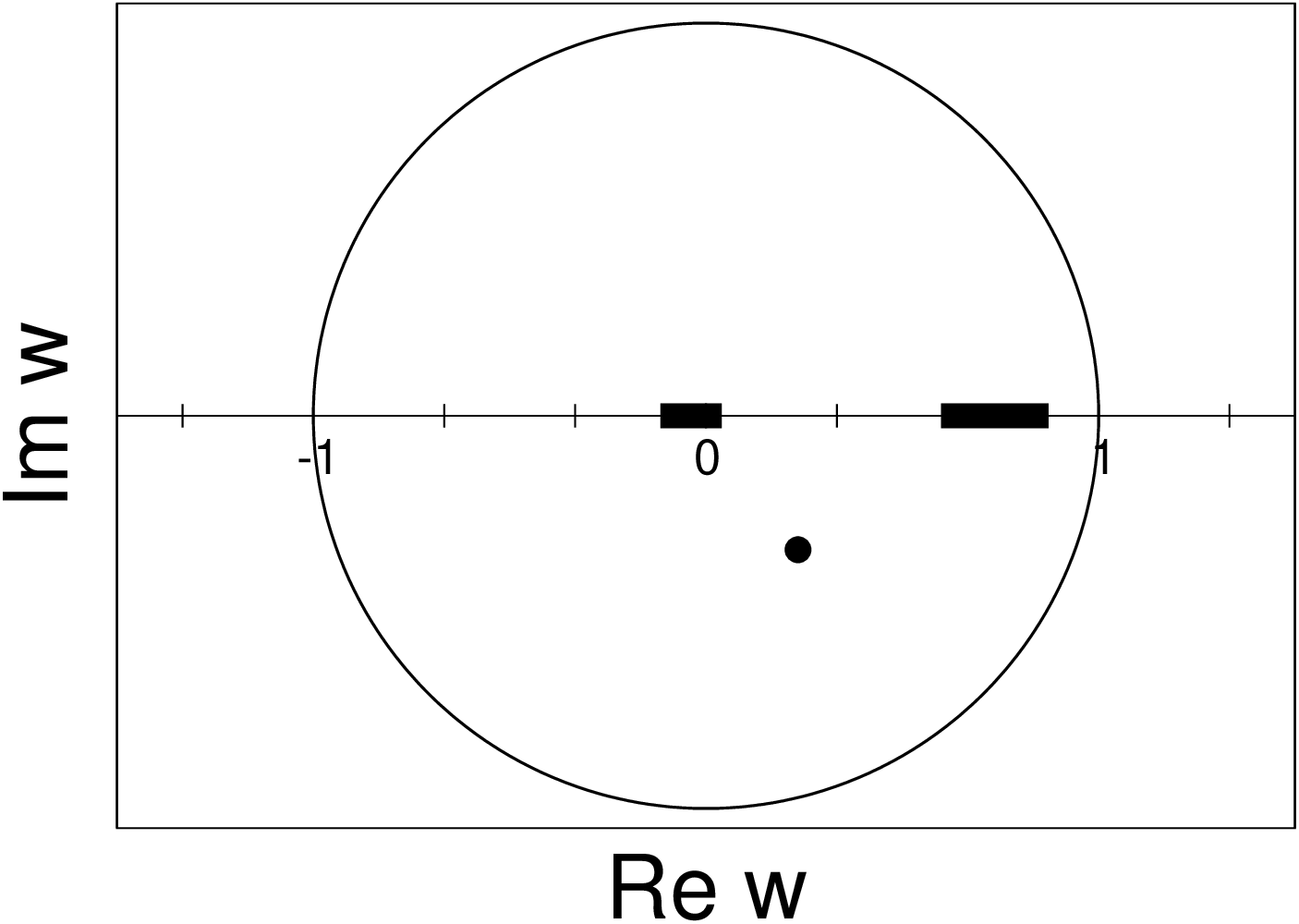}\hspace{0.5cm}\includegraphics[width=4.5cm]{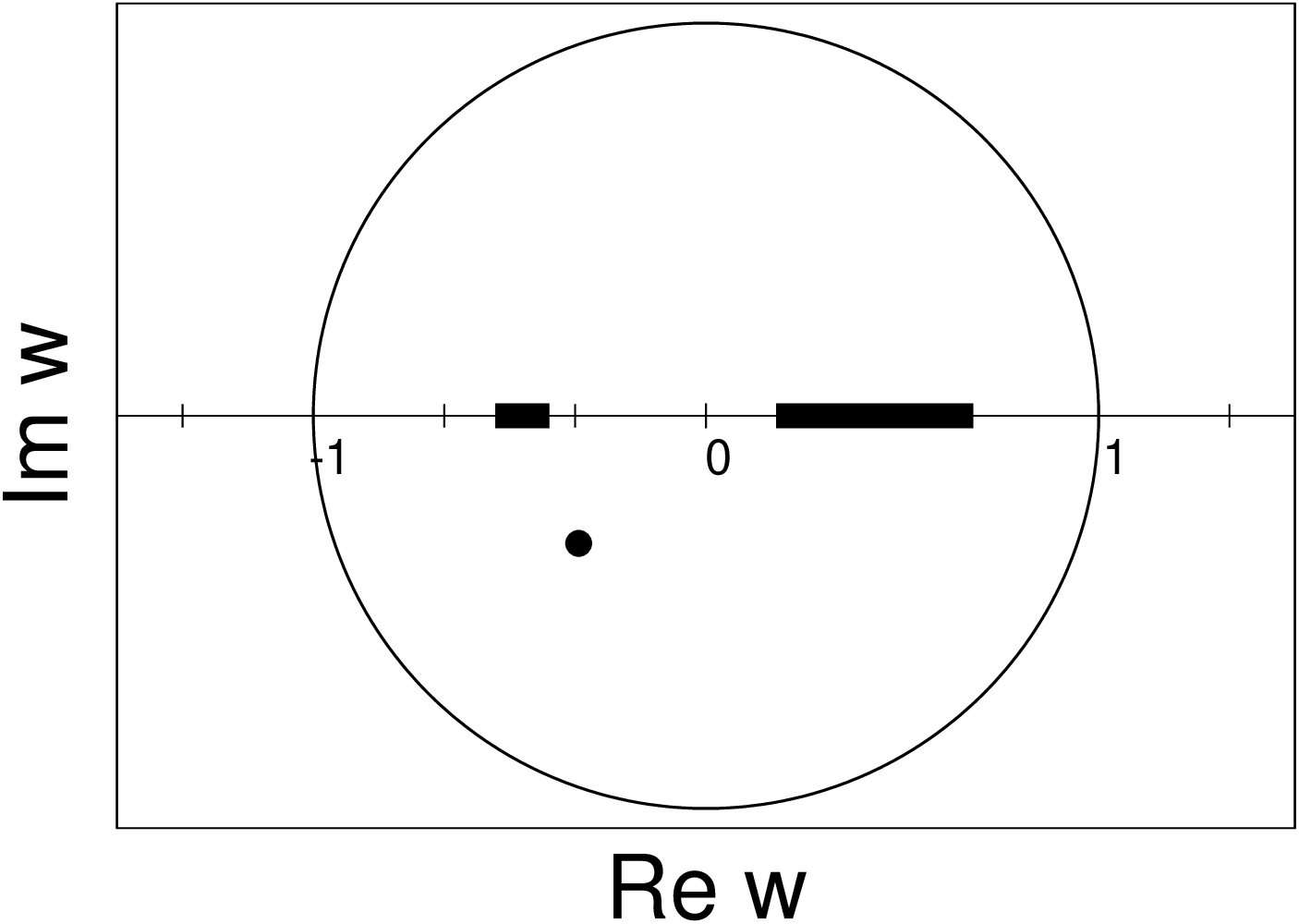}\hspace{0.5cm}\includegraphics[width=4.5cm]{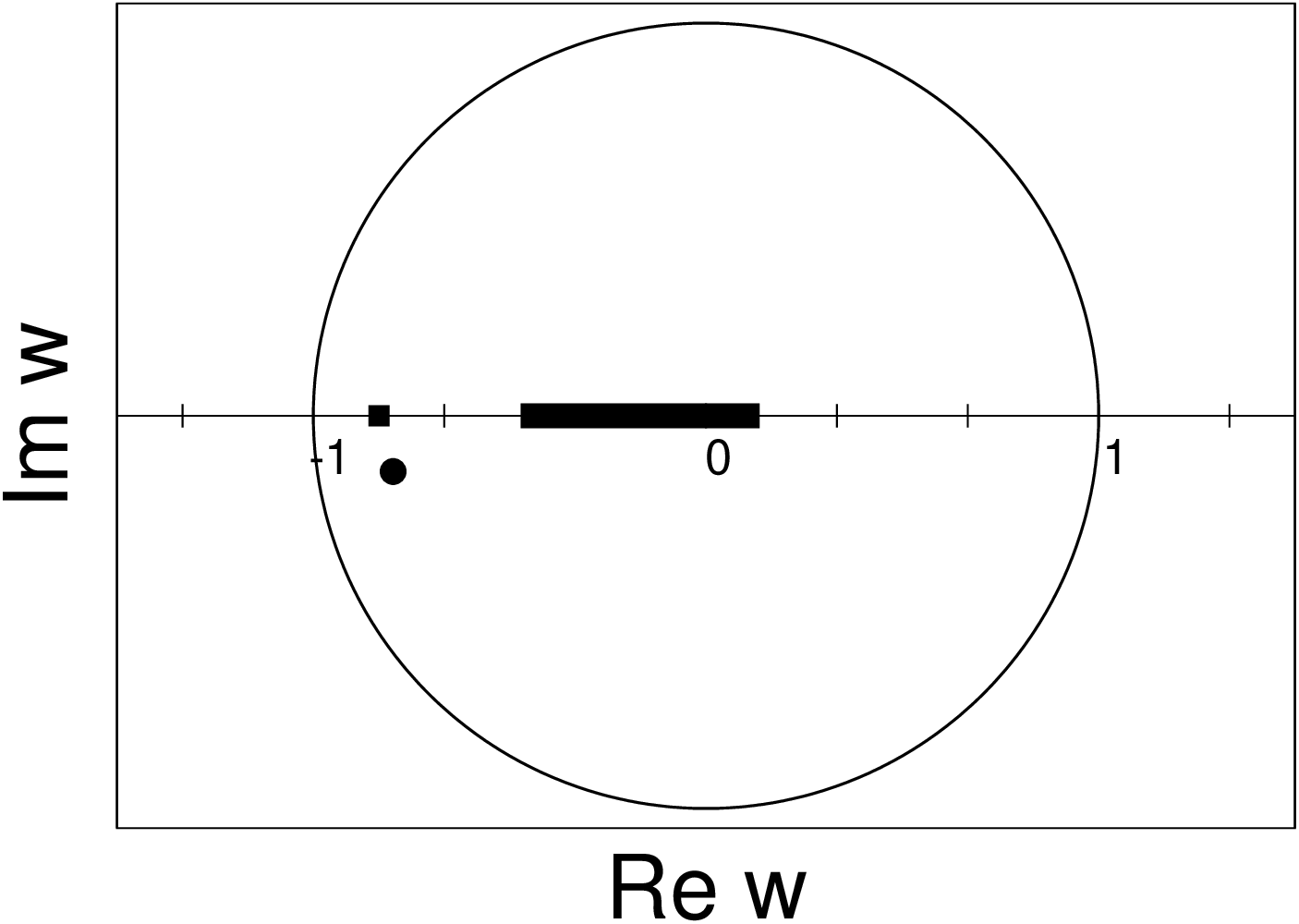}
\end{center}
\caption{The disk $|w|<1$ in the complex plane  $w=w(s,\alpha)$ defined in (\ref{wa}), for $\alpha=0.36$ (left), $\alpha=1$ (center)  and $\alpha=4$ (right). The thick segments indicate the regions where experimental data are available from $K_{e4}$ decay \cite{NA48}-\cite{Pislak} and  the process $\pi N\to\pi\pi N$ (cf. the compilation of data made in \cite{PY 2005}), respectively; the circle shows  the $\sigma$ pole  on the second Riemann sheet from \cite{CCL}.}\end{figure}

\subsection{Alternative procedure for ghost removal}\label{CM}
The singularity at $s=0$ of the phase space factor $\rho(s)$  in (\ref{t00}) can be alternatively eliminated  if
the term $i \rho(s)$ is replaced by a function which is analytic in the $s$-plane cut along $s\ge 4 M_\pi^2$ and has the imaginary part equal to $\rho(s)$ on the upper edge of the cut. In the context of effective range approximation, Chew and Mandelstam \cite{ChMa} defined such a function, vanishing at threshold,  by a once subtracted dispersion relation. For convenience, we  consider the loop function  of ChPT, $\bar{J}(s, M_\pi^2)$, written as
 \be\label{J}
\bar{J}(s, M_\pi^2) =\frac{2}{\pi}+\frac{\rho(s)}{\pi} \ln\left[\frac{\rho(s)-1}{1 + \rho(s)}\right], 
\ee
 which vanishes at the origin, $J(0, M_\pi^2)=0$, and satisfies the relation
\be\label{ImJ}
\mbox{Im}\, \bar{J}(s+i\epsilon, M_\pi^2)= \rho(s), \quad\quad s\ge 4 M_\pi^2.
\ee
If we define the function  $\psi_1(s)$ by:
\be\label{t001}
t_0^0(s)= \frac{1}{\psi_1(s) - \bar{J}(s, M_\pi^2)},
\ee 
the unitarity relation (\ref{unit}) and  Eq. (\ref{ImJ}) show that  $\psi_1(s)$ is real for  $4 M_\pi^2\le s < 4 M_K^2$, where it is  related to the phase shift $\delta_0^0$ by
\be\label{delta001}
 \psi_1(s)= \rho(s)\, \mbox{cot}\delta_0^0(s)+ \mbox{Re}\,\bar{J}(s, M_\pi^2)\,.
\ee  
 The reality property implies also that  $\psi_1(s)$ is analytic in the $s$-plane cut for $s\le 0$ and $s \ge 4 M_K^2$,  except for the pole  at $s=s_A$, and can be expanded as
\be\label{psi1}
\psi_1(s)= \frac{M_\pi^2}{s-s_A}\left[B_0+B_1\, w(s,\alpha )+B_2 \,w(s,\alpha )^2+\ldots \right],
\ee 
in powers of the  variable (\ref{wa}). We remark that the compensating term $2 s_A/M_\pi^2\sqrt{s}$ appearing in  (\ref{psiYgen}) is no longer 
 necessary in (\ref{psi1}), since the  function $\bar{J}(s, M_\pi^2)$ is by definition regular at $s=0$.

 \subsection{$S$-matrix factorization}\label{DT}

 Other parametrizations of $t_0^0(s)$ are obtained by including some information about  its behaviour  near the $K\bar{K}$ threshold.  We do this by expressing the $S$-matrix element
\be \label{S00}
S_0^0(s) =  1+ 2 i \rho(s) t_0^0(s)
\ee
 as a 
  product
\be\label{DalitzTuan}
S_0^0(s)= S_{\rm rest}(s) S_{f_0}(s),
\ee
where each factor satisfies elastic  unitarity ($|S_{\rm rest}(s)|=| S_{f_0}(s)|=1$) below the $K\bar K$ threshold.
 The multiplication of the two $S$-matrices amounts to the following addition rule for the corresponding amplitudes:
\be\label{t}
t_0^0(s)= t_{\rm rest}(s)+  t_{f_0}(s) +  2 i\rho(s)\ t_{\rm rest}(s)\ t_{f_0}(s),
\ee
where
\be\label{tS}
t_{\rm rest}(s)= \frac{S_{\rm rest}(s)-1}{2 i\rho(s)}, \quad\quad
t_{f_0}(s)=\frac{S_{f_0}(s)-1}{2 i\rho(s)}.
\ee
 Crossing symmetry  implemented by Roy equations \cite{CGL} implies that the expansion of the partial
  wave amplitude around $s=0$ starts with
\be \label{ts0}
t_0^0(s)= t_0 + t_1 s + t_2 s^{3/2}+ O(s^2)\,,
\ee
where $t_0$ is nonzero.
In order to cancel the singularity of the factor $\rho(s)$ at $s=0$ in  (\ref{t}), either 
 $t_{\rm rest}(s)$ or  $t_{f_0}(s)$ must vanish at $s=0$, but not both (since, cf. (\ref{ts0}), 
 the full amplitude does not have a zero there). We choose to set $t_{f_0}(0)=0$, taking   for this amplitude the expression\be\label{tf0}
 t_{f_0}(s)= \frac{k_1 s } {\kappa -s- k_1 s \,\bar{J}(s, M_\pi^2) -(k_2+k_3 s) \bar{J}(s, M_K^2)},\ee
where $\bar{J}(s, M_\pi^2)$ is defined in (\ref{J}) and $\bar{J}(s, M_K^2)$ is obtained  replacing   $M_\pi$ in $\rho(s)$ by $M_K$. We note that by taking
\be\label{kappaCohen}
 \kappa = 1.01, \quad  k_1 = 0.08, \quad k_2 = -1.09, \quad k_3 = 1.16,
\ee
the modulus of the corresponding $S$-matrix, $S_{f_0}(s)$, is close, in the range $2 M_K < \sqrt{s} < 1.16$ GeV,  to the elasticity $\eta_0^0(s)$ measured in \cite{Cohen}, while for
 \be\label{BES1u}
\kappa =  1.15, \quad  k_1 = 0.11, \quad k_2 =  0.39, \quad k_3 = 0.03,\ee
\be\label{BES1l}
\kappa =  1.41, \quad  k_1 = 0.24, \quad k_2 = -0.73, \quad k_3 = 1.72, \ee
  $|S_{f_0}(s)|$ is close,  in the same range, to the upper/lower edges of the  band of the elasticity $\eta_0^0$  extracted from  the decay $J/\psi\rightarrow\phi\pi\pi$ \cite{BESf0}. 

For our purpose, the specific form adopted for  $t_{f_0}(s)$  is not a limitation, since the total amplitude contains the additional term $t_{\rm rest}(s)$.    Elastic unitarity, $|S_{\rm rest}(s)|=1$, implies that $t_{\rm rest}(s)$ can be written, for instance, as  
\be\label{trest}
  t_{\rm rest}(s)=\frac{1}{\psi_{\rm rest}(s) -i \rho(s)}\,,  
  \ee
with $\psi_{\rm rest}$  analytic in the $s$-plane cut along $s\le 0$ and $s \ge 4M_K^2$, except for a pole  at  $s=s_1$, where $t_{\rm rest}(s_1)=0$ (from (\ref{t}) and (\ref{tf0}) it follows that $s_1$ is close to the Adler zero $s_A$).  Therefore, we can write $\psi_{\rm rest}(s)$ as:
\be\label{psirest}
\psi_{\rm rest}(s)= \frac{M_\pi^2}{s-s_1} \left[ \frac{2 s_1}{M_\pi \sqrt{s}}+ B_0+B_1 w(s,\alpha ) +B_2 w(s,\alpha )^2+\ldots \right],
\ee
where  $w(s,\alpha )$ is defined in (\ref{wa}). Alternatively, we can use for  $t_{\rm rest}(s)$ an expression similar to (\ref{t001}), involving the function $\bar{J}(s,M_\pi^2)$.

 Other admissible parametrizations are obtained if we assume that   $t_{\rm rest}(s)$ is almost regular near $s=4M_K^2$. Since the next branch point, at $s=4 M_\eta^2$,   is known to have a weak effect,  we can neglect at low energies the right hand cut of  $\psi_{\rm rest}(s)$,  and expand it as
\be\label{psirest1}
\psi_{\rm rest}(s)= \frac{M_\pi^2}{s-s_1} \left[ \frac{2 s_1}{M_\pi \sqrt{s}}+ B_0+B_1 w_1(s,\alpha ) +B_2 w_1(s,\alpha )^2+\ldots \right],
\ee
where the variable 
\be\label{w1a}
 w_1(s,\alpha )=\frac{ \sqrt{s} - \alpha }{\sqrt{s}+ \alpha},\quad\quad \alpha>0,
\ee 
maps the $s$-plane cut only for $s\le 0$ onto the disk $|w_1|<1$ of the complex plane $w_1=w_1(s,\alpha)$. 
 
 The expressions given in this subsection are examples of possible analytic parametrizations of the amplitude at low energies. Following an idea of Dalitz and Tuan \cite{DaTu}, in some phenomenological analyses \cite{BuggZou} the individual $S$-matrices in the product (\ref{DalitzTuan}) are associated to specific resonances. However, in our work we use the factorization  (\ref{DalitzTuan})  only for mathematical purposes: by isolating a factor with a rapid variation near the  $K\bar K$ threshold, we expect a better convergence for the expansion of the remaining part,  $t_{\rm rest}(s)$. This part, which is fixed by the low energy data, contributes to both the elasticity $\eta_0^0(s)= |S_0^0(s)|$ and the phase shift $\delta_0^0$ above the $K\bar K$ threshold. So, the behaviour of $t_0^0$  above this point is  left free in our fits. 
\section{Fits of the data from $K_{e4}$ decay}

We consider first the data on the difference  $\delta_0^0-\delta_1^1$ measured below 0.4 GeV from $K_{e4}$ decay \cite{NA48}-\cite{Pislak}. The $P$-wave phase shift $\delta_1^1$
is known with precision in this energy range \cite{CGL, PY 2005}, allowing an accurate extraction of  $\delta_0^0$.
 As shown recently \cite{Gasser}, the phase shift measured in $K_{e4}$ decay differs from the pure strong phase shift $\delta_0^0(s)$ by an isospin  correction overlooked so far, accounting for the differences between the masses of the charged and neutral mesons, and between the quark masses  $m_u$ and $m_d$.  The correction evaluated in  ChPT to one-loop reads \cite{Gasser}:
\be\label{isocorr}
\Delta[\delta_0^0(s)]= \frac{1}{32 \pi F_0^2} \{(4 \Delta_\pi +s)\rho (s)+(s-M_{\pi^0}^2) \left(1+\frac{3}{2 R}\right) \rho_0(s) -(2 s-M_\pi^2) \rho(s)\},
\ee
where $\rho (s)$ is defined in (\ref{unit}) and
\be 
 \Delta_\pi =M_\pi^2 -M_{\pi^0}^2, \quad \rho_0(s)=\sqrt{1-4 M_{\pi^0}^2/s}, \quad R=\frac{m_s-\hat{m}}{m_d-m_u}\,,\quad\hat{m}=(m_u+m_d)/2\,.
\ee
With the estimate R=37 $\pm$ 4 given in  \cite{Gasser}, the correction $\Delta[\delta_0^0(s)]$ amounts to a fraction of a degree in the whole experimental range.  This correction  was subtracted  from the  phase shift derived from $K_{e4}$ data, in order to obtain the pure strong phase shift $\delta_0^0(s)$.

  The total number of points from the $K_{e4}$ experiments is 21 (5 points from  \cite{Rosselet}, 6 from  \cite{Pislak} and 10 from  \cite{NA48}). As in \cite{GMPY}, we increased the experimental error on the last point in \cite{Pislak}   by 50\%.  For the 10 data from the NA48/2 experiment we used the covariance matrix  published recently in \cite{NA48}. 
We fitted these data with the parametrizations described in section 2.

\begin{figure}\label{FitsKe4}\begin{center}
  \includegraphics[width=7.5cm]{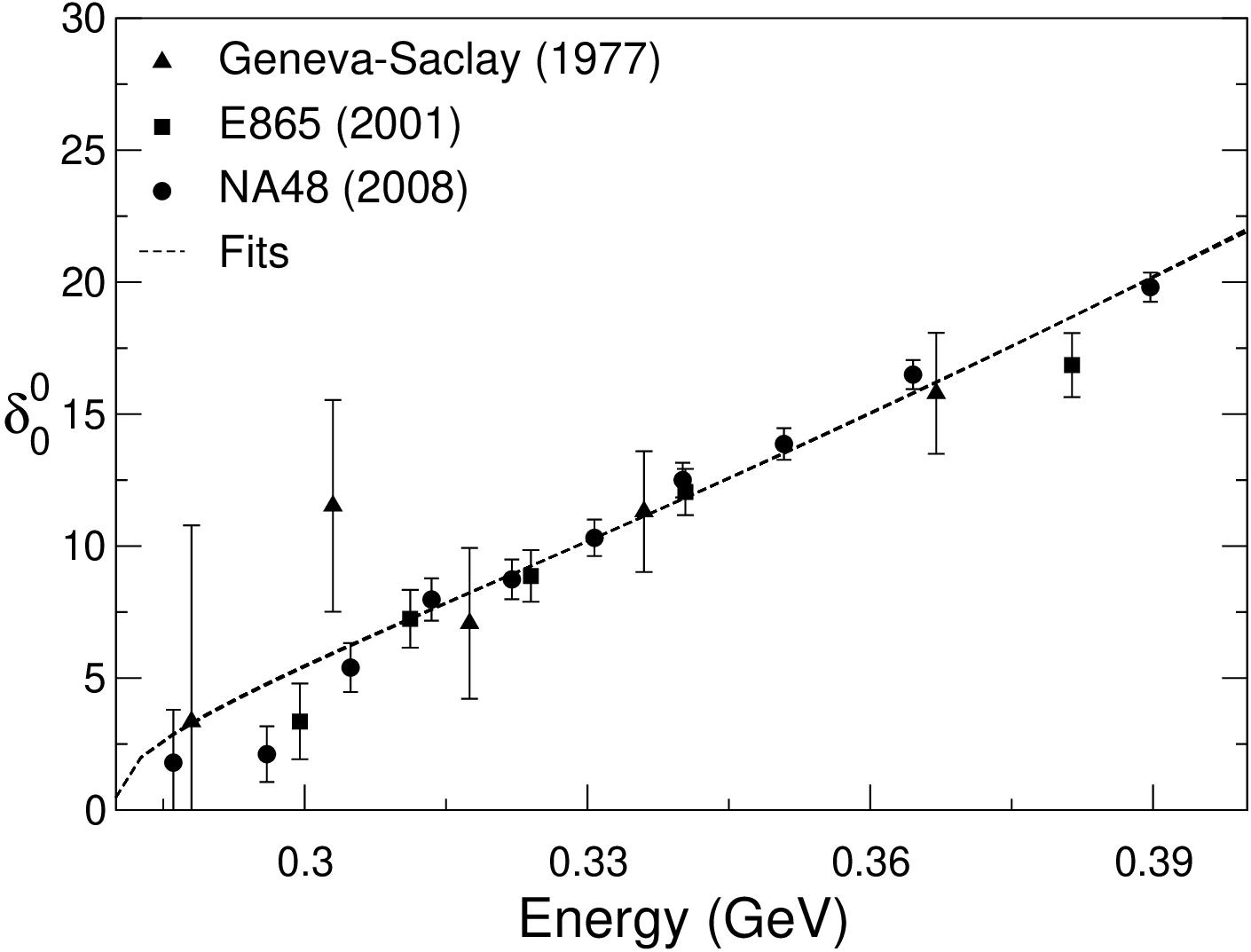}\hspace{0.4cm} \includegraphics[width=7.5cm]{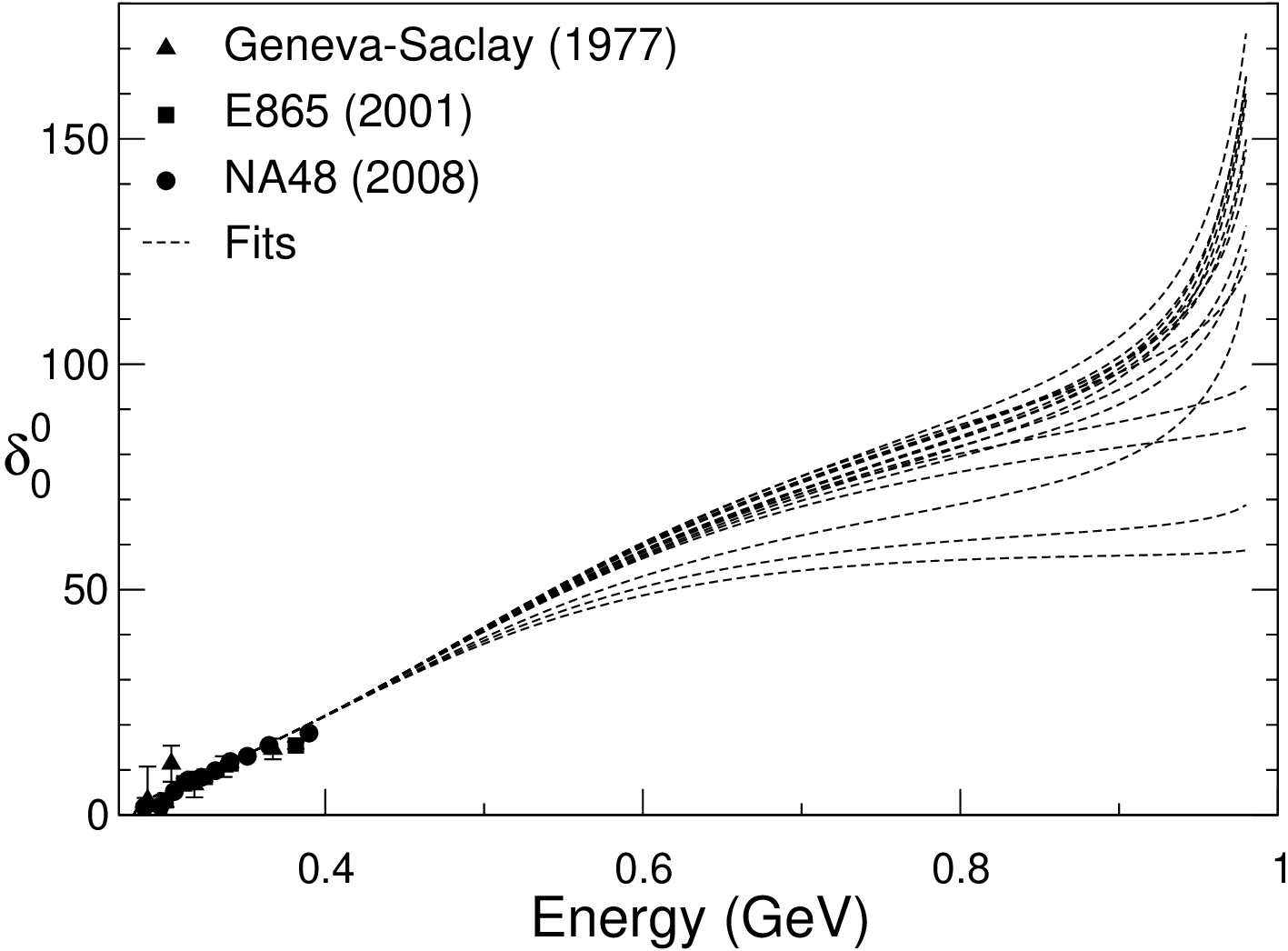}
  \caption{Left: phase shift  $\delta_0^0$ derived from $K_{e4}$ decay, fitted with the 16 parametrizations described in the text. Right: extrapolation of the fits above the experimental range.}\end{center}
\end{figure}

 In our analysis, the positive number $\alpha$, specifying  the conformal variables (\ref{wa}) and (\ref{w1a}), together with the parameters $\kappa$ and $k_i$ appearing in (\ref{tf0}), represent the input which defines an admissible class.  To account for the uncertainty in the position of the Adler zero, $s_A$ was varied between $0.4 M_\pi^2$ and $0.6 M_\pi^2$.
In each admissible class, the coefficients  $B_i$ of the expansion in  powers of the conformal variable are free. They are determined by fitting the low energy data.

We investigated a large class of combination of input parameters, from which we retained
 16 admissible parametrizations: the first three  are based on Eqs. (\ref{t00}) and (\ref{psiYgen}), with $\alpha=1$, $\alpha=0.36$ and  $\alpha=4$, respectively. The value  $\alpha=0.36$ is special, since, as seen in Fig. 1, it maps the  experimental range relevant in $K_{e4}$ decay onto a symmetric interval around the origin $w=0$. According to general theorems \cite{CutDeo}, this variable
 gives the best approximation of data in the experimental range. On the other hand, for $\alpha$ greater than 1 (for instance, $\alpha=4$) the variable $w(s,\alpha)$  maps the region close to the branch point $s=4 M_K^2$ near the origin of the $w$-plane (see right panel of Fig. 1). As we shall see,  the expansion  (\ref{psiYgen}) is then  more suitable at higher energies.

The next three parametrizations are  based on Eqs. (\ref{t001}) and (\ref{psi1}), with the same choices $\alpha=1$, $\alpha=0.36$ and  $\alpha=4$. The difference from the previous fits is the way of eliminating the singularity at $s=0$ of the phase space $\rho(s)$  in the denominator of $t_0^0(s)$.

  The remaining ten
 parametrizations are based on the $S$-matrix factorization discusssed in subsection \ref{DT}.  In three of them  we take the parameters $\kappa$ and $k_i$ from (\ref{BES1u})  and  the expansion (\ref{psirest1}), with  $\alpha=0.36$, $\alpha=1$ and  $\alpha=4$, and in the next two we  use  the same parameters $\kappa$ and $k_i$  and  the expansion (\ref{psirest}), with  $\alpha=0.36$ and $\alpha=1$. In other two cases we use the parameters $\kappa$ and $k_i$ from (\ref{BES1l}) and  the expansion (\ref{psirest1}) with  $\alpha=0.36$ and $\alpha=1$,
  while in the last three  parametrizations we take the parameters $\kappa$ and $k_i$ from  (\ref{kappaCohen}), using either  the expansion (\ref{psirest})  with  $\alpha=1$ and $\alpha=0.36$, or the expansion (\ref{psirest1}) with $\alpha=0.36$. 

 We obtained good fits of the 21 experimental points with 2 free parameters, $B_0$ and $B_1$, in the expansion in powers of the conformal variables. The values of $\chi^2$ are very similar for all the fits, although the parametrizations  are quite different. The values of $\chi^2$ and the optimal parameters  are given in Table 1, for $s_A$ (or $s_1$) fixed at $0.5 M_\pi^2$ (note that if $s_1=0.5 M_\pi^2$,  the position $s_A$ of the Adler zero, resulting from the fits, is slightly different: for the last 10 fits given in Table 1, $s_A$ varied between  $0.42 M_\pi^2$ and $0.47 M_\pi^2$). The values of  $\chi^2$ decrease by about 0.4 if we take into account the theoretical uncertainty associated to the isospin correction \cite{Gasser}. For simplicity we indicate only the central values of the parameters, omitting the statistical errors.

\begin{table}\label{TableKe4}\begin{center}\vskip0.3cm
\begin{tabular}{rcccccl}
Nr. &  $\chi^2$ & $B_0$ & $B_1$ & $a_0^0$ & $b_0^0$ & $\sqrt{s_\sigma}\, (\mbox{MeV}) $\\\hline
1.&  21.7&    7.5&  -15.1&   0.216& 0.278&  459 + 259 i \\
2.&  21.5&   14.6&  -12.4&   0.214& 0.282&  445 + 259 i \\ 
3.&  21.9&  -16.4&  -37.1&   0.217& 0.275&  473 + 261 i \\
4.&  20.9&    7.7&  -20.2&   0.212& 0.287&  412 + 237 i \\
5.&  20.6&   17.2&  -16.5&   0.210& 0.292&  401 + 231 i \\
6.&  21.2&  -35.0&  -60.3&   0.214& 0.284&  422 + 246 i \\
7.&  21.5&   14.6&    -14.8& 0.214& 0.281&  443 + 262 i \\
8.&  21.7&   5.8&   -18.8&   0.215& 0.278&  455 + 261 i \\
9.&  21.8&  -25.3&  -47.9&   0.216& 0.276&  465 + 260 i \\
10.& 21.6&   15.2&  -12.8&   0.215& 0.280&   451 + 264 i \\
11.& 21.8&   7.8&  -15.5&    0.216& 0.277&  466 + 264 i \\
12.& 21.5&   15.0&  -15.3&   0.214& 0.281&  448+ 264 i \\
13.& 21.7&   5.9&  -19.5&    0.216& 0.277&  459 + 262 i \\
14.& 21.8&   7.8&  -15.5&    0.216& 0.277&  465 + 263 i \\
15.& 21.6&   15.1&  -12.8&   0.215& 0.280&  450 + 263 i \\
16.& 21.4&   14.5&  -14.8&   0.214& 0.282&  443 + 261 i \\\hline
\end{tabular}\caption{Results of the fits of the data from $K_{e4}$ decay \cite{NA48}-\cite{Pislak}, using the 16 parametrizations  described in the text.}\end{center}\end{table}

The quality of the fits is seen in Fig. 2. Although the fits are almost indistinguishable in the experimental range, they exhibit large differences when extrapolated to higher energies. This illustrates the well-known phenomenon of instability of analytic extrapolation \cite{Ciulli}. We note that the lowest curves in the right panel of Fig. 2  correspond to the fits 4, 5 and 6 in Table 1, obtained with  the parametrization (\ref{t001})-(\ref{psi1}). In particular, the fits no. 4 and 5, corresponding to the choices $\alpha=1$  and $\alpha=0.36$ in (\ref{psi1}), exhibit a plateau at low values of $\delta_0^0(s)$.  The  increase of the phase shift required by the high energy data (see below) is  obtained, for instance, with the choice $\alpha=4$ in the expansion (\ref{psiYgen}), or by using parametrizations based on the $S$-matrix factorization described in subsection 2.3.

In Table 1 we give for each fit the central values of $a_0^0$ and $b_0^0$ and the position $s_\sigma$ of the $\sigma$ pole on the second Riemann sheet, obtained by analytic extrapolation to the threshold and into the complex plane (as shown in \cite{CCL},  $s_\sigma$ is the solution of the equation $S(s_\sigma)=0$ on the first sheet).

Taking the average of the 16 admissible values for $m_\sigma=\sqrt{s_\sigma}= M_\sigma -i\Gamma_\sigma/2$,  we obtain:
\be\label{sigmaKe4}
M_\sigma=447 \pm 7\, \mbox{(stat)} \,\,^{+25}_{-46} \,\mbox{(syst)} \,\, \mbox{MeV},\quad \Gamma_\sigma/2= 258 \pm 6\, \mbox{(stat)}\, \,^{+10}_{-26}\, \mbox{(syst)}\,\, \mbox{MeV},
\ee
where the systematic error is defined, as in \cite{GMPY}, such as to cover all the admissible fits (the uncertainty in the Adler zero produces a small error, of about 4 MeV in  $M_\sigma$ and 3 MeV in $\Gamma_\sigma/2$).

\begin{figure}\label{sigmamass}\begin{center}
  \includegraphics[width=12cm]{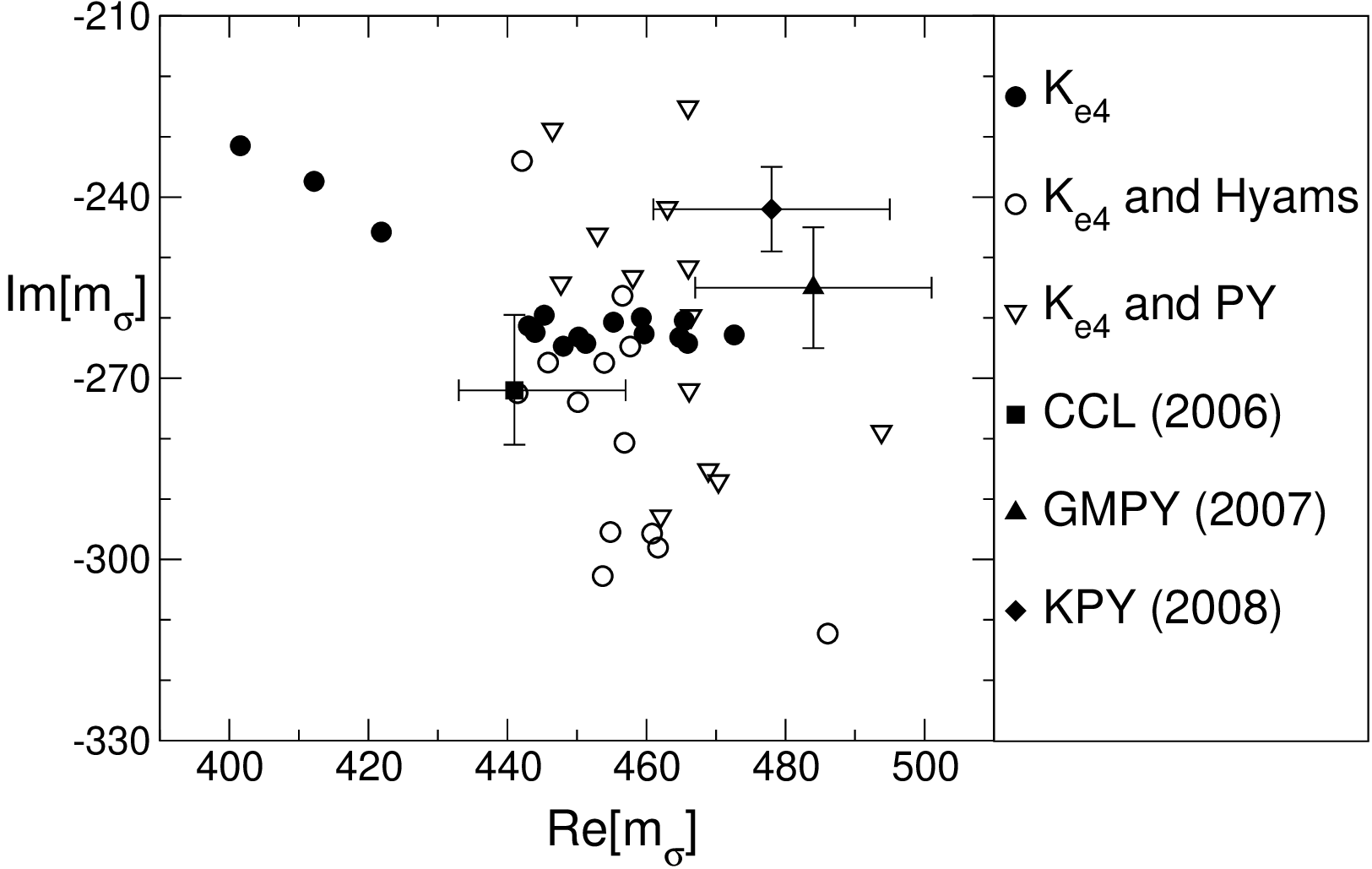}
  \caption{Positions of the $\sigma$ pole obtained by the analytic extrapolation of the parametrizations used for fitting various sets of data, compared with   Refs. \cite{CCL}, \cite{GMPY} and \cite{KPY 2007} (from the last reference we show the value  obtained with  isospin corrected $K_{e4}$ data).}
\end{center}\end{figure}

The pole positions given in Table 1 are
shown in Fig. \ref{sigmamass}, together with the results reported in  \cite{CCL}, \cite{GMPY} and \cite{KPY 2007}.  Note that the three isolated points, with small values of $M_\sigma$ and  $\Gamma_\sigma$, correspond to the fits 4, 5 and 6 in Table 1, which are based on the parametrization (\ref{t001})-(\ref{psi1}). As we mentioned, they lead to a bad behaviour of the phase shift at higher energies, in spite of the fact that they provide very good fits of the data from $K_{e4}$ decay.  From Table 1 it is seen that these parametrizations (especially 4 and 5) give low values   for    $a_0^0$ and large values  for $b_0^0$.

In this Section we obtained  good fits of the phase shift measured from $K_{e4}$ decay. However,  the extrapolation of the phase shift above the experimental region is not acceptable for many of them. Therefore, we  can not take
 the results of this Section as final. In particular,  the narrow range  spanned by most of the widths $\Gamma_\sigma$  in Fig.  \ref{sigmamass} may signal a  bias. In the next Section we shall improve the description of $t_0^0(s)$  by including data on the phase shift at higher energies.

\section{Inclusion of high energy data}

\begin{table}\label{TableHyams}\begin{center}\vskip0.3cm
\begin{tabular}{rcccccccr}
Nr. &  $\chi^2$ &   $\chi^2_{Ke4}$ &  $B_0$ & $B_1$ & $B_2$ & $a_0^0$ & $b_0^0$ & $\sqrt{s_\sigma}\, (\mbox{MeV}) $\\\hline
1.&  37.7& 24.3  &  7.8&  -23.5& -20.6 &  0.233&  0.261     & 486 + 312 i \\
2.&  32.9& 22.8  &   -20.3&  -61.5&  -23.9 & 0.226& 0.271   & 462 + 298  i \\ 
3.&  32.6& 22.6  &  -37.3&  -84.6&  -29.6 &  0.225&  0.272   & 461 + 296 i \\
4.&  32.7&  22.2 &    1.3&  -52.3&  -39.9 & 0.222  &  0.271  & 458 + 265 i \\
5.&  33.9& 21.8 &   -52.8&  -105.8& -26.0  &  0.207  & 0.283   & 442 + 234 i \\
6.&  32.4& 22.1 &    9.3&  -12.0& -  &  0.220   & 0.274   & 457 + 281 i \\
7.&  33.7&22.9  &  -2.2&    -20.3& - & 0.228  & 0.272   & 454 + 303 i \\
8.&  33.9&  21.8  &  13.1  & -12.2  & -  &   0.215    & 0.278  & 454 + 267 i \\
9.&  38.0& 21.8  &5.2 &  -19.9 & -  &   0.213   &  0.278  & 456 + 256 i \\
10.&  32.1&21.5  & 18.8  &-12.9  & - &  0.216   &  0.281  & 441 + 272 i \\
11.& 31.9& 21.6  & 12.4  & -13.9  & -  &  0.216    &  0.279  &  446 + 267 i \\
12.& 32.1& 21.9  &  7.4&  -15.5&-  &   0.219  &  0.276  &  450 + 274 i \\
13.& 32.7& 22.6  &  -9.5&  -29.3& -  &  0.225   &  0.272  &  455 + 295 i \\\hline
\end{tabular}\caption{Results of the fits of the 40 data points of set I, using the 13 parametrizations described in the text.}\end{center}\end{table}

The difference $\delta_0^0(s)-\delta_0^2(s)$   is measured at  $s=M_K^2$ from the decay $K\to\pi\pi$ \cite{Aloisio}. However, in  this case the radiative corrections are very large  \cite{Cirigliano} and the extraction of the strong phase shift  $\delta_0^0$ is still uncertain \cite{Colangelo}. For this reason, we shall not use as input in our analysis this datum.

 Experimental data  at higher energies are available from  the  $\pi N\to\pi\pi N$ process \cite{Hyams, Proto}.
 We considered  two sets of data below the $K\bar K$ threshold:
\begin{itemize}
\item set I, which consists from 40 data points: 21  from  $K_{e4}$ decay \cite{NA48}-\cite{Rosselet} and  19  from the CERN-Munich experiment \cite{Hyams};
\item  set II, which consists from 32 data points: 21 from  $K_{e4}$ decay \cite{NA48}-\cite{Rosselet}  and a collection of 11 data points from $\pi N\to\pi\pi N$ \cite{Proto}, given in Eq. (2.13) of \cite{PY 2005}.
\end{itemize}

 As in the previous section, we investigated a large number of parametrizations, but rejected many of them since they gave bad fits. For instance, the choice $\alpha=0.36$ (or other values  $\alpha<1$) in the expansions (\ref{psiYgen}) and (\ref{psi1}) was not admissible, leading to high values of $\chi^2$ (such parametrizations can not exhibit the rapid increase of the phase shift above 900 MeV). Also, the parametrization (\ref{psiY}), considered in \cite{GMPY}, proved to be not acceptable: with 3 free parameters, $\mu_0$, $B_0$ and  $B_1$, we obtained  $\chi^2=45.6$ for the 40 points of set I, and  $\chi^2=34.6$ for the 32 points of set II. We recall that  expressions  which display the energy where the phase shift passes through $\pi/2$ are often used
 for fitting narrow resonances.  However, they are not suitable for broad resonances like $\sigma$.

We finally  retained 13 admissible parametrizations:  the first three  are based on Eqs. (\ref{t00}) and (\ref{psiYgen}), with $\alpha=1$, $\alpha=4$ and  $\alpha=6$, respectively.
The next two  are  based on Eqs. (\ref{t001}) and (\ref{psi1}), with the  choices $\alpha=1$ and  $\alpha=4$.
  The remaining eight
 parametrizations are based on the $S$-matrix factorization discusssed in subsection \ref{DT}: in four cases we use  the values $\kappa$ and $k_i$ from (\ref{kappaCohen}), and either the expansion (\ref{psirest}) with  $\alpha=1$, $\alpha=4$ and $\alpha=0.5$, respectively, or the expansion (\ref{psirest1})  with  $\alpha=1$. In the next three cases we use  the parameters $\kappa$ and $k_i$ from (\ref{BES1u}), and either the expansion  (\ref{psirest}) with  $\alpha=0.2$, or  the expansion  (\ref{psirest1}), with  $\alpha=0.5$ and $\alpha=1$. Finally, in  the last parametrization  we take the values of $\kappa$ and $k_i$ from (\ref{BES1l}) and the expansion (\ref{psirest1})  with $\alpha=4$.

  For the first five parametrizations we used 3 nonzero coefficients,  $B_0$, $B_1$ and $B_2$, in the expansion in powers of the conformal variables,  and for the last eight we obtained good fits with 2 nonzero coefficients,  $B_0$ and  $B_1$. As in the previous Section,  we took into account the uncertainty in the position of the Adler zero, by allowing in each case $s_A$ to vary between 0.4 $M_\pi^2$ and 0.6 $M_\pi^2$.

\begin{figure}\label{delta00HyamsPY}
  \includegraphics[width=7.5cm]{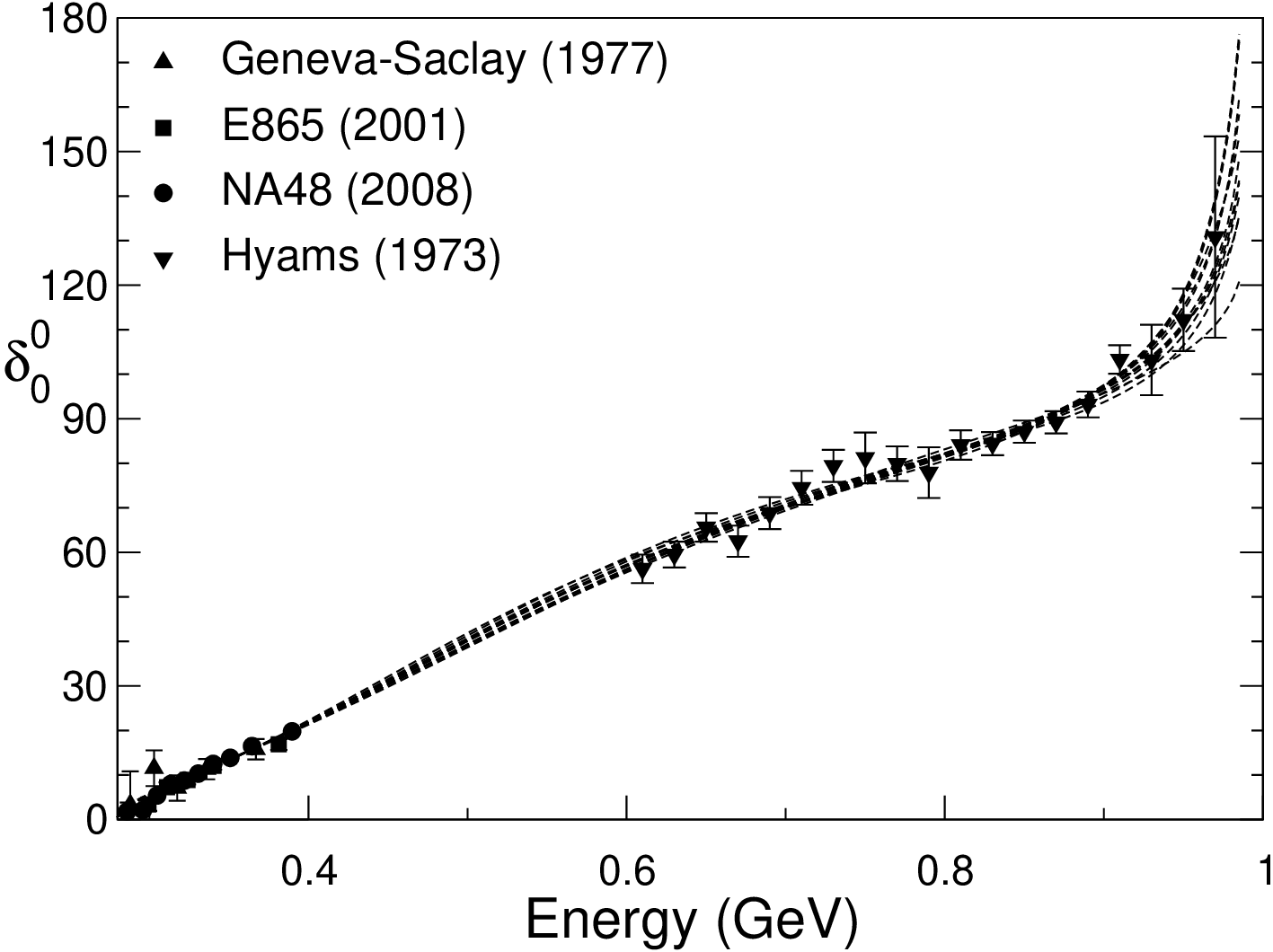}\hspace{0.5cm}
\includegraphics[width=7.5cm]{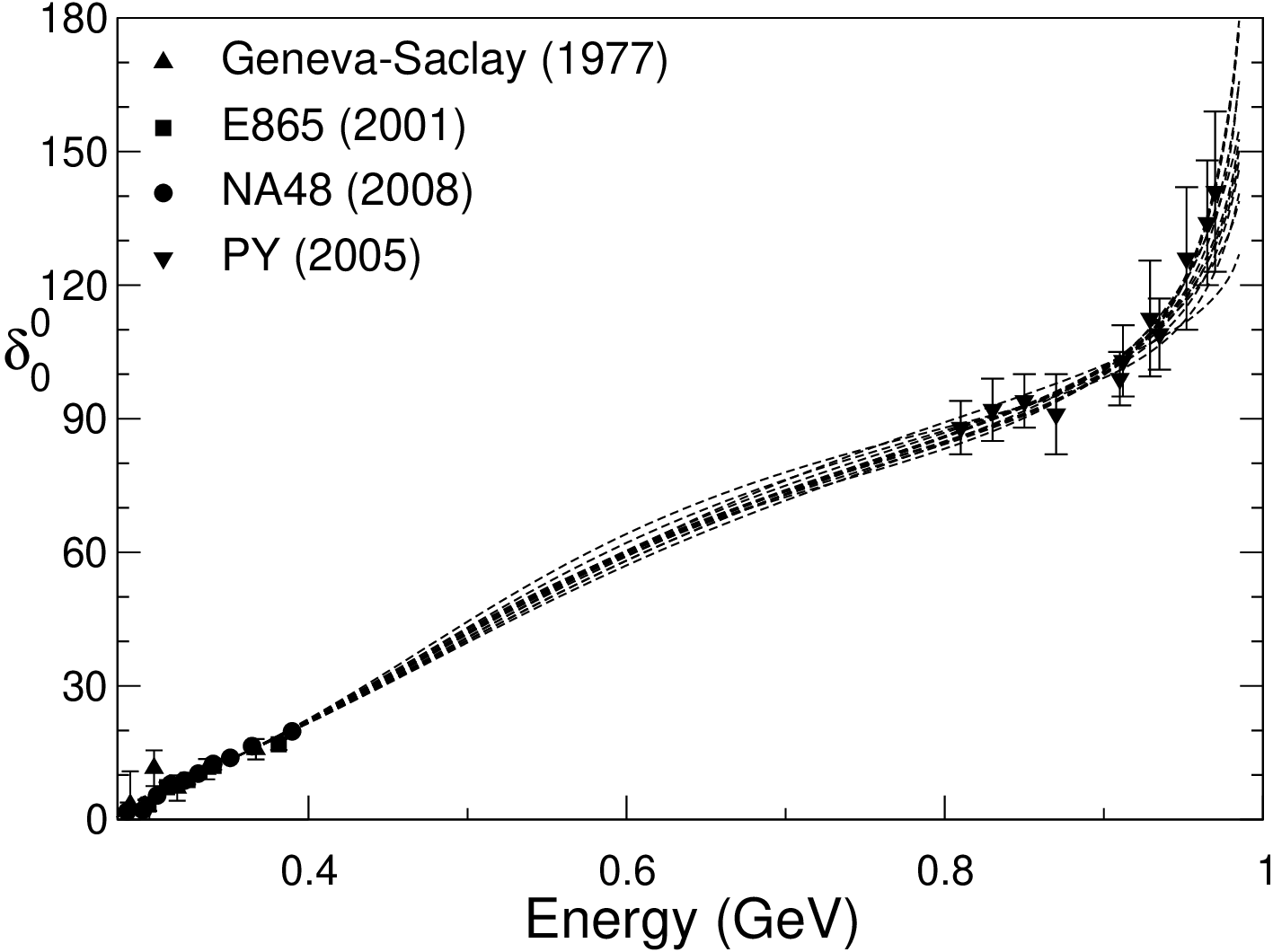} \caption{Left: fits of the data in set I ($K_{e4}$ data \cite{NA48}-\cite{Pislak}  plus the CERN-Munich data \cite{Hyams} below the $K\bar K$ threshold). Right:  fits of the data in set II ($K_{e4}$ data  plus a selection of data from $\pi N\to\pi\pi N$, given in  \cite{PY 2005}).}
\end{figure}

 The results of the fits (for $s_A$ or $s_1$ fixed at $0.5 M_\pi^2$) are presented in Tables 2 and 3, for the sets I and II, respectively.  For completeness, we show also in column 3  the contribution to the $\chi^2$ of the 21 points from $K_{e4}$ decay.  By comparing Tables 2 and 3 with Table 1 we notice that the description of the $K_{e4}$ data,  measured by their contribution to the total $\chi^2$, is now slightly worse than in the fits restricted to the data from $K_{e4}$ decay:  constraining the behaviour at high energies leads to a small deterioration of the description of the low energy data. The quality of the fits is shown in Fig. \ref{delta00HyamsPY}, and in  Fig. 5 we show an expanded view of the energy region covered by $K_{e4}$ decay. As expected from the previous discussion,  the various parametrizations are no longer indistinguishable, as were those obtained from fitting only the $K_{e4}$ data in Fig. 2.

\begin{table}\label{TablePY}\begin{center}\vskip0.3cm
\begin{tabular}{rcccccccr}
Nr. &  $\chi^2$ &   $\chi^2_{Ke4}$ &  $B_0$ & $B_1$ & $B_2$ & $a_0^0$ & $b_0^0$ & $\sqrt{s_\sigma}\, (\mbox{MeV}) $\\\hline
1.&  28.6& 22.8  & 5.4&  -29.0& -20.9 &  0.225&  0.267     & 494 + 279 i \\
2.&  24.8& 22.4  &   -26.6&  -71.8&  -27.1 & 0.223& 0.272   & 470 + 287  i \\ 
3.&  24.4& 22.3  &  -45.9&  -97.4&  -33.0 &  0.223&  0.273   & 469 + 285 i \\
4.&  25.1 & 22.0 &  -0.3&  -56.6&  -41.3 & 0.218  &  0.273  & 466 + 251 i \\
5.&  24.6& 22.2 &   -56.6&  -111.1& -26.9  &  0.205  & 0.284   & 446 + 229 i \\
6.&  23.9& 21.8 &    7.4&  -16.3& -  &  0.215   & 0.277  & 466 + 259 i \\
7.&  24.3&22.5  &  -6.2&    -25.3& - & 0.224  & 0.272   & 462 + 293 i \\
8.&  24.7&  22.1  &  12.5&   -16.9& -  &   0.208    & 0.281  & 463 + 242 i \\
9.&  26.6& 22.9  &  1.0&   -28.8& -  &   0.204   &  0.282  & 465 + 225 i \\
10.& 23.2&21.6  &  20.1&  -16.8& - &  0.210  &  0.284  & 448 + 254 i \\
11.& 23.2& 21.8  &  11.6&  -18.7& -  &  0.209    &  0.283  &  453 + 246 i \\
12.& 23.1& 21.8  &  4.8&  -20.8&-  &   0.213  &  0.279  &  458 + 253 i \\
13.& 23.5& 22.0  &  -21.1&  -43.3& -  &  0.219   &  0.274  &  466 + 272 i \\\hline
\end{tabular}\caption{Results of the fits of the 32 data points of set II, using the 13 parametrizations described in the text.}\end{center}\end{table}

Taking the average of the values of the threshold parameters $a_0^0$ and $b_0^0$  given in Tables 2 and 3,  weighted with the ratio $\chi^2/{\rm N_{dof}}$,  we obtain, for the two sets:
\bea\label{abHyams}
&&a_0^0=0.220 \pm 0.005\, \mbox{(stat)} \pm 0.013 \,\mbox{(syst)} \pm 0.003 (s_A) \quad \mbox{(I)} \nonumber \\
&&a_0^0=0.215 \pm 0.005\, \mbox{(stat)} \pm 0.011 \,\mbox{(syst)} \pm 0.003 (s_A)\quad \mbox{(II)} 
\eea
and
\bea\label{abPY}
&&b_0^0= 0.275 \pm 0.006 \mbox{(stat)}^{+0.009}_{-0.014}\mbox{(syst)}\pm 0.004 \,(s_A)\quad \mbox{(I)} \nonumber \\
&&b_0^0= 0.277 \pm 0.006 \mbox{(stat)}^{+0.007}_{-0.010}\mbox{(syst)}\pm 0.004 \,(s_A) \quad \mbox{(II)}.
\eea
If we combine these determinations we obtain the values
\be\label{abfinal}
a_0^0=0.218 \pm 0.014, \quad\quad \quad b_0^0=0.276 \pm  0.013,\ee
which are fully consistent  with the results  obtained from ChPT and low energy theorems for $\pi\pi$ scattering \cite{CGL}, quoted in (\ref{abCGL}).

Before discussing the results for the $\sigma$ pole, let us make a few comments on the phase shift $\delta_0^0(s)$ shown in Fig. 4. It was advocated in some papers, for instance \cite{KPY 2007}, that the phase shift of the isoscalar $S$-wave exhibits a "hump" at energies around 800 MeV,  before starting the rapid increase near the $K\bar K$ threshold.   As seen from  Fig. 4, no hump is seen in the fits of the data in the set I, while  a weak hump appears only in a few fits of the data in set II. This proves that the hump seems to be an artefact of special parametrizations used for fitting the data (in particular, the expression (\ref{psiY}) displays a pronounced hump if a small number of coefficients $B_i$ is kept in the expansion). As discussed in \cite{HLAzore}, this shape is in conflict with the forward dispersion relation for the $\pi\pi$ amplitude of isospin $I=0$.

It is of interest to calculate with our parametrizations the value of  $\delta_0^0$  at  $\sqrt{s}=M_K$.  We recall that  the phase shift difference  $\delta_0^0(s)-\delta_0^2(s)$ at this energy can be extracted  from the decay $K\to\pi\pi$. However, as we mentioned in section 4, in this case the isospin breaking corrections are large and the extraction of the strong phase shift  $\delta_0^0$ is still unclear. For this reason  we did not use this information as input in our fits. Using the parameters given in Tables 2 and 3 and taking the averages of the admissible values in the two sets we obtain:
\bea\label{deltamk}
&&\delta_0^0(M_K^2)= 38.9^\circ \pm 0.6^\circ\, \mbox{(stat)} ^{+ 1.7^\circ}_{-1.4^\circ}\,\mbox{(syst)} \quad \quad \mbox{(I)} \nonumber\\
&&\delta_0^0(M_K^2)= 40.4^\circ \pm 0.8^\circ\, \mbox{(stat)} ^{+ 2.6^\circ}_{-1.8^\circ}\,\mbox{(syst)} \quad \quad \mbox{(II)}.
\eea
In \cite{GMPY} the authors used as input in their fits the value  $\delta_0^0(M_K^2)=48.7 \pm 4.9^\circ$, which is significantly larger than the output values given in (\ref{deltamk}).

\begin{figure}[htb]\label{delta00HyamsPYLow}\vspace{0.5cm}\begin{center}
  \includegraphics[width=7.5cm]{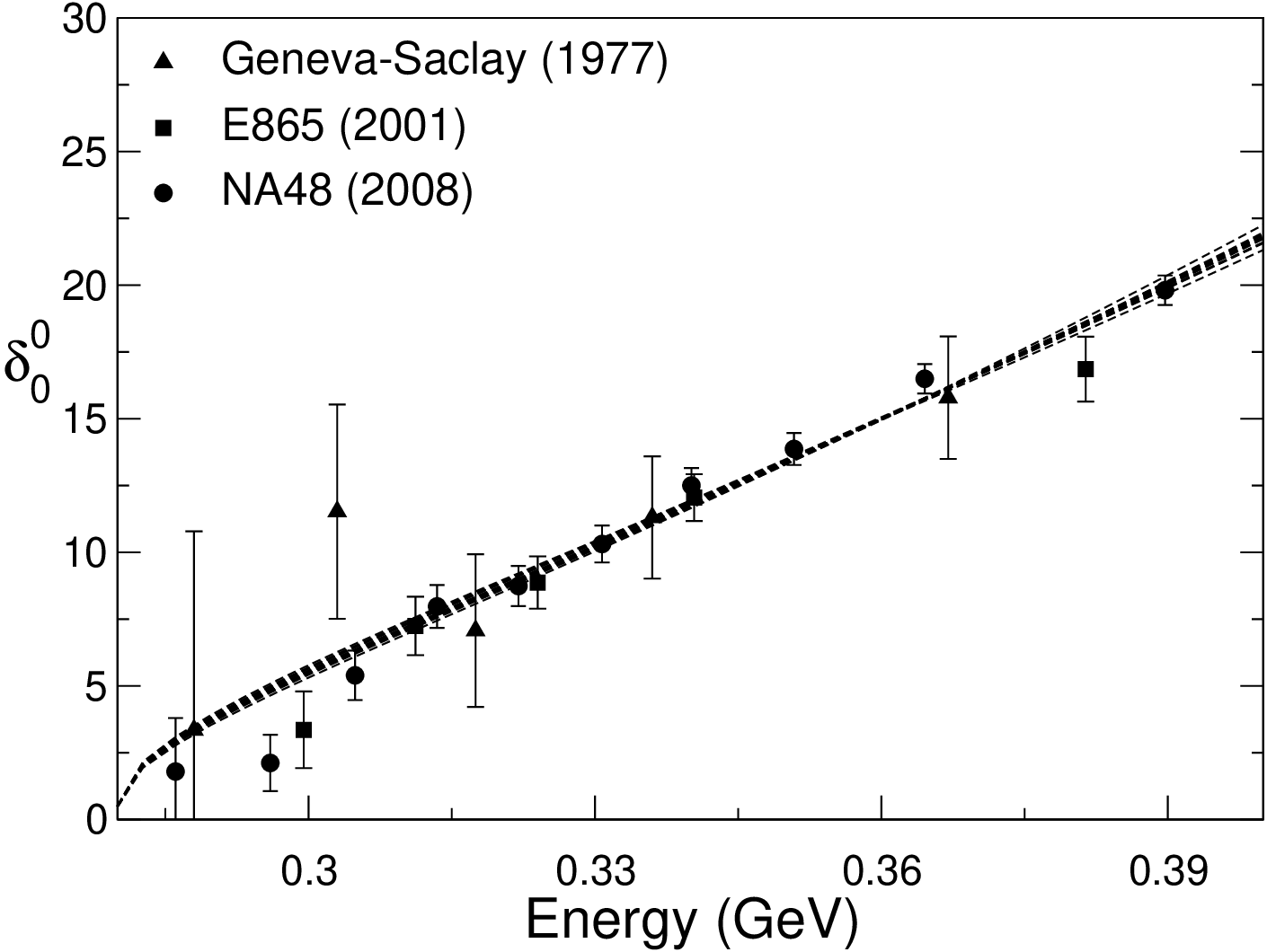}\hspace{0.5cm}
\includegraphics[width=7.5cm]{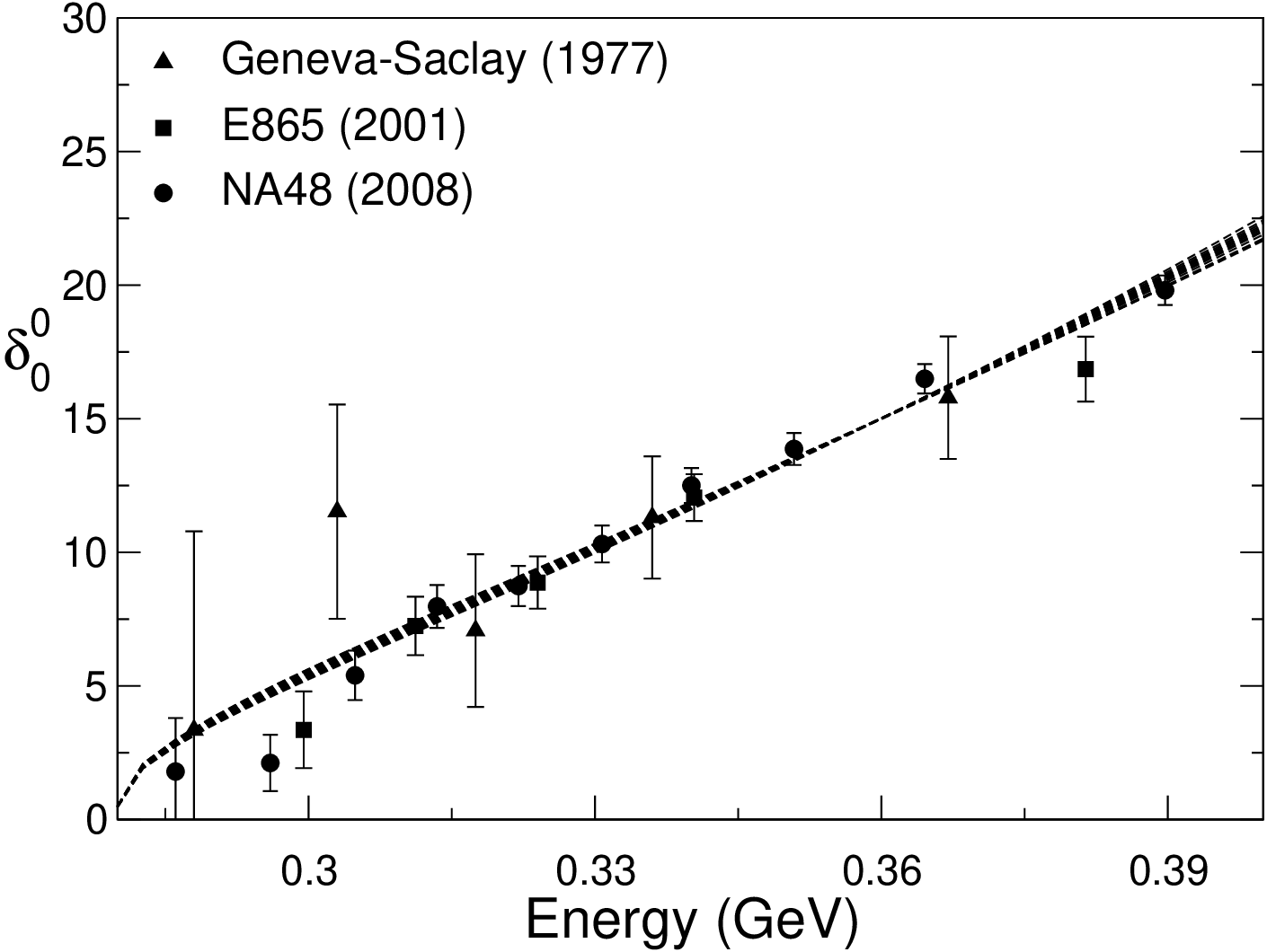} \caption{Expanded view of the low energy  region  from  Fig. \ref{delta00HyamsPY}.}\end{center}
\end{figure}

\begin{figure}\label{eta00}\begin{center}
  \includegraphics[width=8.5cm]{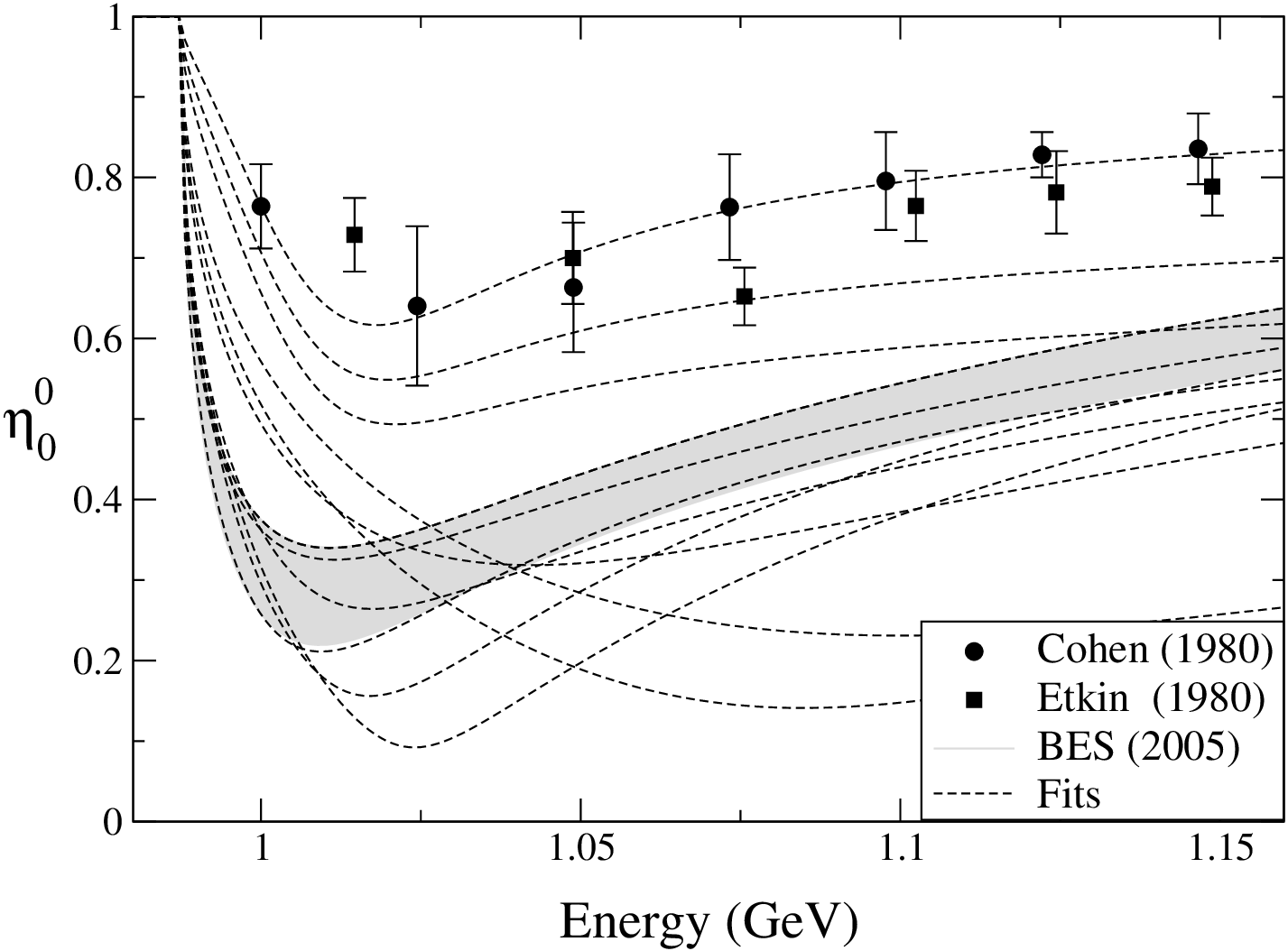}\caption{Elasticity $\eta_0^0$ obtained with the 13 parametrizations of the data in set I (cf. Table 2), compared with  the experimental data from \cite{Cohen}-\cite{BESf0}. }\end{center}
\end{figure}

The phase shift at 0.8 GeV is also of interest, since it is a key input in solving Roy equations \cite{ACGL}.  We recall that in \cite{ACGL} the range is  $\delta_0^0(0.8\, \mbox{GeV} ) = 82.3 \pm 3.4 ^\circ $ was adopted as input, while in  \cite{CGL} the more conservative  choice  $\delta_0^0(0.8\, \mbox{GeV} ) = 82.3 ^{+10^\circ}_{-4 ^\circ} $ was made. From the parametrizations discussed above and the parameters $B_i$ given in Tables 2 and 3, we obtained the average values:
\bea\label{delta08}
&&\delta_0^0(0.8\, \mbox{GeV} )= 81.8^\circ \pm 0.6^\circ\, \mbox{(stat)} \pm 1.3^\circ\,
\mbox{(syst)}\quad \quad \mbox{(I)} \nonumber\\
&&\delta_0^0(0.8\, \mbox{GeV}  )= 85.9^\circ \pm 0.7^\circ\, \mbox{(stat)} ^{+3.3^\circ}_{-2.6^\circ}\, \mbox{(syst)} \quad\quad \quad \mbox{(II)},
\eea
which are consistent with the ranges adopted in  \cite{ACGL} and  \cite{CGL}.

We turn now to the predictions for the $\sigma$ pole  obtained by the analytic continuation of the parametrizations considered above. The  $\sigma$ pole positions given in Tables 2 and 3 are shown in Fig. \ref{sigmamass}. 
 The three isolated points obtained with some fits of the $K_{e4}$ data are no longer allowed, but in the same time the tight correlation  exhibited by the other fits of the $K_{e4}$ data is now softened. This is due to the fact that the description of the $K_{e4}$ data,  measured by their contribution to the total $\chi^2$, is slightly worse than in the fits restricted only to the $K_{e4}$ data, and the various parametrizations are not as indistinguishable at low energies as in Fig. 2.

The averages of the values given in Tables 2 and 3, respectively, weighted with the corresponding $\chi^2/N_{\rm dof}$, give:
\bea\label{sigma}
M_\sigma=455 \pm 6 \mbox{(stat)}^{+31}_{-13}\mbox{(syst)}\,\mbox{MeV},&\Gamma_\sigma/2= 278 \pm 6 \mbox{(stat)}^{+34}_{-43}\mbox{(syst)}\,\mbox{MeV}&{\rm (I)}\nonumber\\
M_\sigma=463 \pm 6\mbox{(stat)}^{+31}_{-17}\mbox{(syst)}\,\mbox{MeV},&\Gamma_\sigma/2= 259 \pm 6 \mbox{(stat)}^{+33}_{-34}\mbox{(syst)}\,\mbox{MeV}& {\rm (II}).
\eea
As above, the systematic errors cover the values in the admissible samples. The uncertainty in the position of the Adler zero $s_A$ has now a smaller effect, of 2 MeV for $M_\sigma$ and 1 MeV for $\Gamma_\sigma/2$. Alternatively, we can define the central values from the optimal fits with the lowest $\chi^2$ (fit no. 11 in Table 2 and fit no. 12 in Table 3). This procedure gives
\bea\label{sigmaopt}
M_\sigma=446 \pm 6 \mbox{(stat)}^{+40}_{-4}\mbox{(syst)}\,\mbox{MeV},&\Gamma_\sigma/2= 267 \pm 6 \mbox{(stat)}^{+44}_{-33}\mbox{(syst)}\,\mbox{MeV}&{\rm (I)}\nonumber\\
M_\sigma=458 \pm 6\mbox{(stat)}^{+36}_{-11}\mbox{(syst)}\,\mbox{MeV},&\Gamma_\sigma/2= 253 \pm 6 \mbox{(stat)}^{+39}_{-28}\mbox{(syst)}\,\mbox{MeV}& {\rm (II}).
\eea
However, this definition is not very sharp since, as seen from Tables 2 and 3, there are several fits with very close values of $\chi^2$,  which lead to different values for $M_\sigma$ and $\Gamma_\sigma$.

  Eqs. (\ref{sigma}) and (\ref{sigmaopt}) represent our final   results for the $\sigma$ pole position, obtained using the data on $K_{e4}$ decay and two, rather complementary, sets of scattering data at higher energies. The differences between them  indicate the sensitivity of the pole location to the behaviour of the phase shift near the $K\bar K$ threshold.

The comparison of (\ref{sigma})  and (\ref{sigmaopt}) with (\ref{CCL}) shows that the analytic extrapolation of experimental data leads to values for the mass and width of $\sigma$ which are consistent with ChPT and Roy equations, but have larger  theoretical uncertainties. The  errors are produced by the well-known  instability of analytic continuation \cite{Ciulli}: functions very close along  a limited part of the boundary may differ drastically  outside the initial range.  We illustrate  this feature in Fig. 6, where we show the elasticity $\eta_0^0$ calculated with the 13 parametrizations fitting the data of set I, given in Table 2, extrapolated above the inelastic threshold. Also, in Fig. 7, we show the real and the imaginary parts of the same parametrizations  along a range covering a part of the left hand cut.  Note that we are calculating now the amplitude on the cuts of the $s$-plane, while, strictly speaking, the expansions in powers of conformal variables converge only at points inside the analyticity domain \cite{Confmap}.
  However,  the expansions are truncated at low orders and are far from the asymptotic regime, so we may view them as effective parametrizations which have a meaning also on the boundary.

  We recall that information about $t_0^0(s)$ along  the left hand cut (rigorously speaking, only on a part of it) is obtained  from crossing symmetry. The dominant contribution is given by the $\rho$ resonance, which does not make a narrow peak. Of course, in the present approach crossing symmetry is not explicitly  implemented.
Fig. 7 shows  the   differences in the values on the left hand cut of amplitudes  which are almost indiscernable in the physical region. Since  the $\sigma$ pole is rather close to the left hand cut, the spread in the positions of the $\sigma$ pole shown in Fig. \ref{sigmamass} is not surprising.

\begin{figure}\label{ReIm}\begin{center}
  \includegraphics[width=7.5cm]{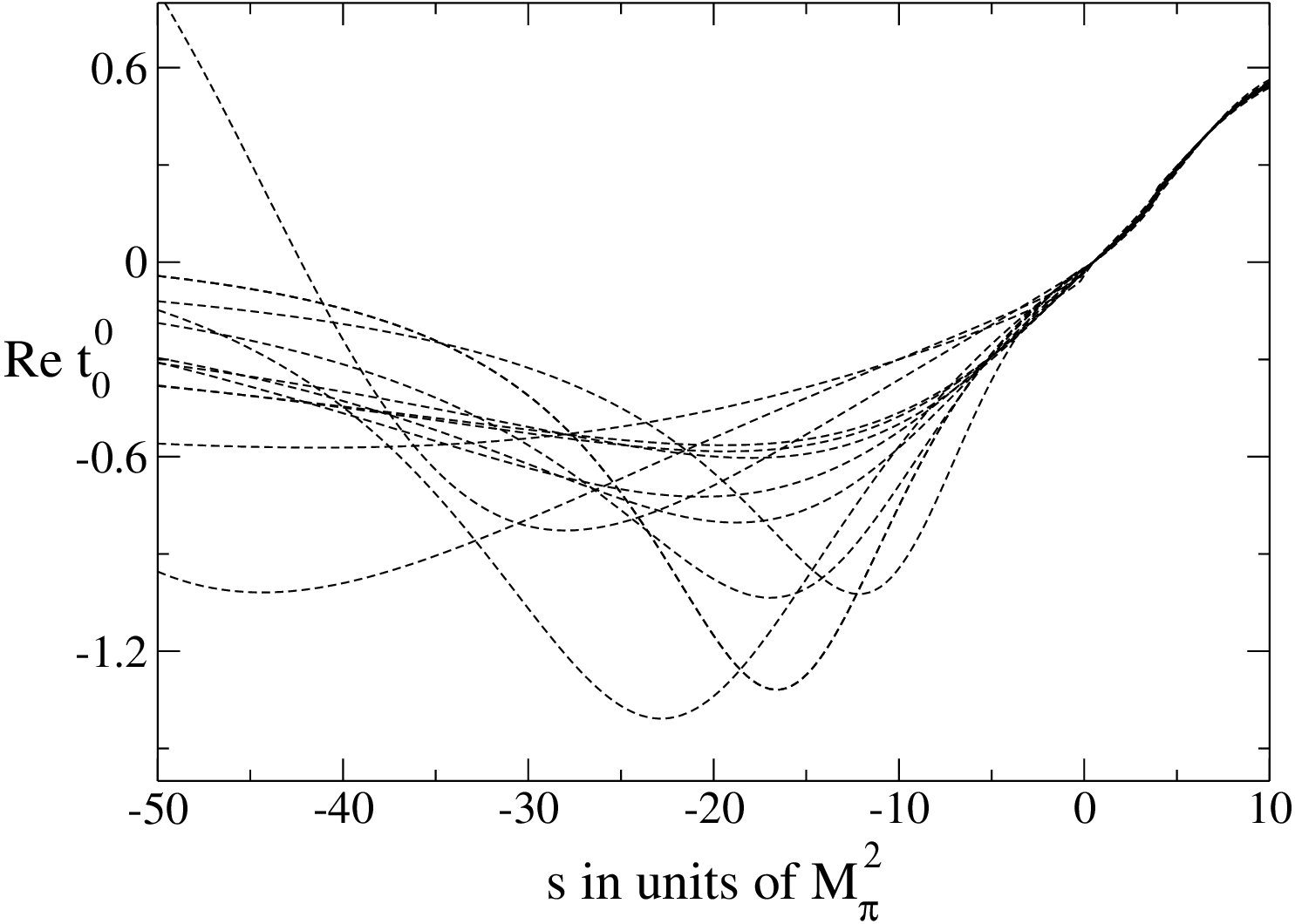}\hspace{0.5cm}
\includegraphics[width=7.5cm]{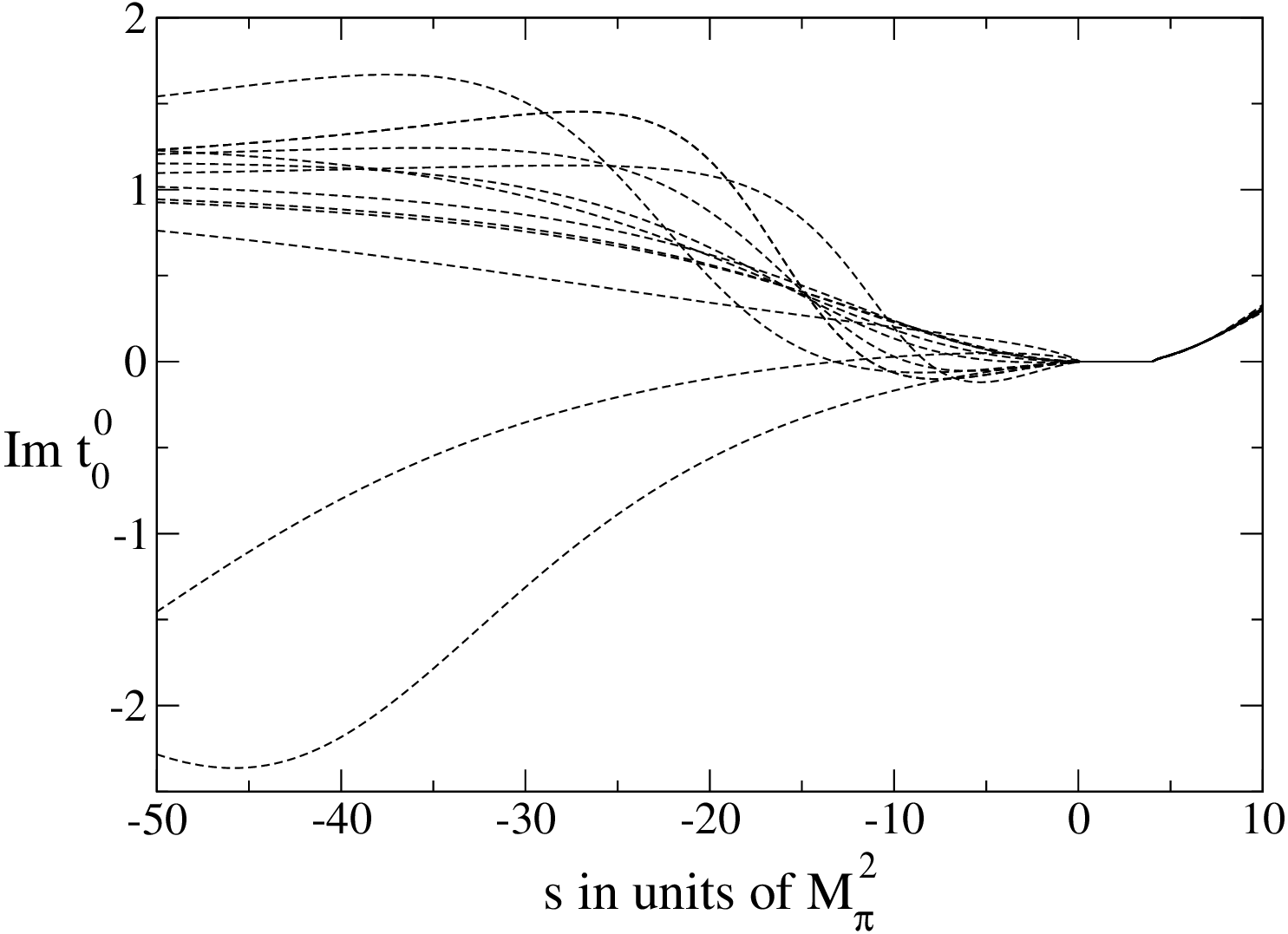} \caption{ $\mbox {Re}\,t_0^0(s)$ (left) and  $\mbox{ Im}\,t_0^0(s)$ (right) obtained by the extrapolation of the 13  parametrizations of the data in set I.}\end{center}
\end{figure}

\section{Conclusions}\label{Conclusions}
The new accurate data on $\pi\pi$ scattering at low energies obtained from $K_{e4}$ decay by the NA48/2 Collaboration \cite{NA48} revived the interest in  finding  the  $\sigma$ pole by the standard method used for narrow resonances, {\it i.e.} by the analytic extrapolation of a suitable parametrization of the partial wave with the quantum numbers of the resonance. 

 In the present work we extended the investigation done in \cite{GMPY}, by using a larger class of analytic functions for the parametrization of the $\pi\pi$  isoscalar $S$-wave at low energies. The purpose was to reduce the theoretical bias and to  provide a more realistic estimate of the systematic uncertainties on the  pole position.

Our analysis shows that, in spite of the remarkable accuracy of the new data obtained from $K_{e4}$ decay \cite{NA48}, the inclusion of data at higher energies is necessary in order to reduce the theoretical bias and to  exclude parametrizations which do not have a suitable behaviour above the experimental range.

 The values (\ref{sigma}) represent our prediction for the  mass and width of $\sigma$, obtained by the  analytic extrapolation of a large number of admissible parametrizations of the isoscalar $S$-wave.  We present separately the results obtained by fitting with the same parametrizations  the two sets of data from the process $\pi N\to \pi\pi N$,  in order to illustrate the sensitivity of the pole position to the behaviour near the $K\bar K$ threshold. We emphasize that, although we used a large number of parametrizations, the admissible sample is still limited. Therefore,  the procedure is not entirely model independent (for a parametrization-free method for the detection of resonances from error-affected data given along a finite range see \cite{CCPS}).

 Our results (\ref{sigma}) are consistent with the mass and width of $\sigma$  obtained   from  ChPT and Roy equations, quoted in (\ref{CCL}).
 However, the method employed here has larger theoretical uncertainties due to the phenomenon of  instability of analytic extrapolation from a part on the boundary \cite{Ciulli}: the differences between the various parametrizations  are amplified by the
 extrapolation from the physical region to a distant point in the complex plane. 
 
The extrapolation error can be kept under control by using  additional information about  the physical amplitude, besides the low energy experimental data.  In the method based on Roy equations, this information is provided mainly by crossing symmetry  and low energy theorems for $\pi\pi$ scattering \cite{CGL}. As shown in \cite{CCL}, this tames the instability of the extrapolation to the $\sigma$ pole,  leading to the small errors quoted in (\ref{CCL}) (for a detailed discussion see also \cite{HLScadron}). We conclude that Roy equations provide the most precise determination of $\sigma$ from $\pi\pi$ elastic scattering.

\vskip0.5cm\noindent
{\bf\large Acknowledgments:}  I thank   F. J. Yndur\'ain for an interesting correspondence which stimulated the present work, and  H. Leutwyler for many useful comments and suggestions.   This work was supported  
by the  Program CEEX of Romanian ANCS under Contract Nr.2-CEx06-11-92.

\end{document}